\documentclass[twocolumn,showpacs,preprintnumbers,amsmath,amssymb]{revtex4}

\usepackage{graphicx}% Include figure file
\usepackage{dcolumn}% Align table columns on decimal point
\usepackage{bm}% bold math
\usepackage{enumerate}
\usepackage[latin1]{inputenc}
\newcommand{\comment}[1]{}

\begin{document}
\setlength{\unitlength}{0.7\textwidth}
%\openup 2\jot
\title{Extreme Lagrangian acceleration in confined turbulent flow}
\author{ Benjamin Kadoch$^{1}$, Wouter J.T.  Bos$^{1,2}$, Kai Schneider$^{1}$} 
%\author{Benjamin Kadoch, Wouter Bos, Kai Schneider}
%\markboth{B. Kadoch, W.J.T. Bos and K. Schneider}{Extreme Lagrangian acceleration in confined turbulent flow}
%
\affiliation{$^1$ Laboratoire de Mod\'elisation et Simulation Num\'erique en M\'ecanique et G\'enie des Proc\'ed\'es (MSNM-GP),
CNRS and Universit\'es d'Aix-Marseille, 38, rue F. Joliot-Curie,\\
13451 Marseille Cedex 20, France\\
$^2$ Laboratoire de M\'ecanique des Fluides et d'Acoustique - UMR CNRS 5509, Ecole Centrale de Lyon - Universit\'e Claude Bernard Lyon 1 - INSA de Lyon, 69134 Ecully Cedex, France} 

%\nodate{}
%
%----------------------------------------------------------------------------
\begin{abstract}
A Lagrangian study of two-dimensional turbulence for two different geometries, a periodic and a confined circular geometry, is presented to investigate the influence of solid boundaries on the Lagrangian dynamics. It is found that the Lagrangian acceleration is even more intermittent in the confined domain than in the periodic domain. The flatness of the Lagrangian acceleration as a function of the radius shows that the influence of the wall on the Lagrangian dynamics becomes negligible in the center of the domain and it also reveals that the wall is responsible for the increased intermittency. The transition in the Lagrangian statistics between this region, not directly influenced by the walls, and a critical radius which defines a Lagrangian boundary layer, is shown to be very sharp with a sudden increase of the acceleration flatness from about $5$ to about $20$.
\end{abstract}


%47.27.E- Turbulence simulation and modeling
%47.27.N- Wall-bounded shear flow turbulence
%47.27.T- Turbulent transport processes
%-------------------------------------------------------------------------------
\maketitle
%\section{Introduction}
The Lagrangian point of view is in many aspects the most natural way to obtain understanding of turbulent transport and mixing. Therefore, for many years, Lagrangian studies have been proposed \cite{Tennekes_1972} but only quite recently, using Direct Numerical Simulation (DNS) \cite{Yeung_1989}, and new experimental techniques  \cite{Mordant_2004,Bodenschatz_2001}, Lagrangian statistics, such as the Lagrangian velocity and acceleration have become fully accessible. For a review on Lagrangian studies in three dimensional turbulence we refer to \cite{Yeung_2002}. Many applications, like the mixing of pollutants in geophysical flows, or the dynamics of plasmas with a strong imposed magnetic field, can be considered within the framework of bidimensional turbulence \cite{Kraichnan_1980,Tabeling_PhysRep}. Moreover, two-dimensional turbulence contains a large range of nonlinearly interacting scales, which is a feature it shares with three-dimensional turbulence. Thanks to its lower dimensionality, two-dimensional turbulence is then a convenient test-bed for a first approach of physical phenomena in three-dimensions.
 In two-dimensions, Lagrangian statistics have been obtained for isotropic turbulence \cite{Elhmaidi_1993,Beta_2003}, in which it was shown that the probability density function (PDF) of the Lagrangian velocity is close to Gaussian and that the coherent structures are responsible for the transport.
 From three-dimensional experiments \cite{Bodenschatz_2001} and DNS \cite{Mordant_2004}, it is known that the Lagrangian acceleration shows a more intermittent behavior than the velocity. Surprisingly, in two-dimensional turbulence, Lagrangian acceleration has not been studied so far. In all previous numerical studies mentioned, periodic boundary conditions allow the fluid elements to freely move in all directions. On the contrary, in the aforementioned experimental works, Lagrangian statistics were obtained in a cylindrically confined flow between two counter-rotating disks, in which the fluid elements are hindered in their motion by the presence of solid boundaries. It can be argued that practically all flows are wall bounded and a fine knowledge of the influence of solid boundaries on the turbulent flow is therefore of major importance. In particular, the investigation of the influence of confinement on Lagrangian statistics is useful for the analysis of experimental results in which the effects due to the walls are usually unknown.

In the present letter we will address this question: what is the influence of solid boundaries on the Lagrangian statistics? The problem is investigated in the framework of two-dimensional turbulence. The influence of boundaries in confined two-dimensional flows was previously studied in several works \cite{Clercx_PRL_2000,Heijst_2006,Schneider_PRL_2005}, focusing merely on Eulerian statistics. In \cite{Clercx_PRL_2000} it was shown that the build up of a boundary layer altered the Eulerian spectral energy density. This boundary layer can be expected to influence the Lagrangian statistics, the investigation of which is the subject of the present work.

%_______________________________________________________________________________ 
%\section{Numerical method}
%
We now describe the method. In order to assess the influence of walls, we consider two distinct geometries: a biperiodic and a circular domain with no-slip boundary conditions.  Two-dimensional incompressible turbulent flow with unit density is considered, governed by the Navier-Stokes equations written in dimensionless form in vorticity-velocity formulation:
\begin{equation}\label{NS}
\frac{\partial \omega}{\partial t} + \vec u \cdot \nabla \omega - \nu \nabla ^2 \omega = - \frac{1}{\eta} \nabla \times (\chi  \vec u)\; ,
\end{equation}
where $\vec u = (u_1,u_2)$ is the velocity, $\omega = \nabla \times \vec u$ is the vorticity, $\nu$ is the kinematic viscosity. The term on the right hand side is a volume penalization term, that is responsible for the boundary conditions \cite{Angot_1999,Schneider_CF_2005}, and which is not present in the periodic case. The mask function $\chi$ is $1$ outside the flow-domain where no-slip walls are to be imposed and $0$ inside the flow, where the Navier-Stokes equations are recovered. The permeability $\eta$ is chosen sufficiently small for given $\nu$ \cite{Schneider_CF_2005} in order to insure the convergence of the volume penalization method. No external forcing is present in equation (\ref{NS}). Typically, numerical investigations of the Lagrangian dynamics are performed in turbulence, forced by a random isotropic stirring to obtain a statistically stationary flow. The choice of a similar forcing in a bounded domain is less trivial. Furthermore, the presence of forcing involves a model, so that for a proper comparison between confined and periodic flow we choose to consider the freely decaying case.

The numerical scheme is based on a classical pseudo-spectral method with resolution $N= 1024^2$, and a semi-implicit time integration with $\Delta t= 5.10^{-5}$ \cite{Schneider_PRL_2005,Schneider_CF_2005}. The Lagrangian quantities are calculated by interpolating the Eulerian quantities and integrated in time using a second order Runge-Kutta scheme. The Lagrangian acceleration is the sum of the gradient of pressure and viscous diffusion $\vec a_L = -\nabla p + \nu \nabla^2 \vec u $. We compute the Lagrangian statistics averaged over $1020$ trajectories, for each geometry. The viscosity is $\nu=10^{-4}$, and the permeability is $\eta=10^{-3}$. For both cases the initial condition corresponds to a Gaussian correlated random field, with an initial enstrophy $Z = \frac{1}{2} \langle \omega^2 \rangle_x = 127$ ($\langle \cdot \rangle_x$ denotes the spatial average), an eddy turn over time $T_e=1/\sqrt{2Z}=0.062$ and a Taylor microscale $\lambda = \sqrt{E/Z}=0.056$, where $E = \frac{1}{2} \langle {\vec u}^2 \rangle_x $ is the initial kinetic energy. For the periodic geometry, the initial Reynolds number is $Re=S\sqrt{E}/\nu=5\cdot10^{4}$, where $S=2\pi$ corresponds to the domain size. For the circular geometry the initial Reynolds number is $Re=2R\sqrt{E}/\nu=4.5\cdot10^{4}$ where $R=2.8$ is the radius of the circle. 

\begin{figure}
 \begin{center}
 \includegraphics[scale=.5]{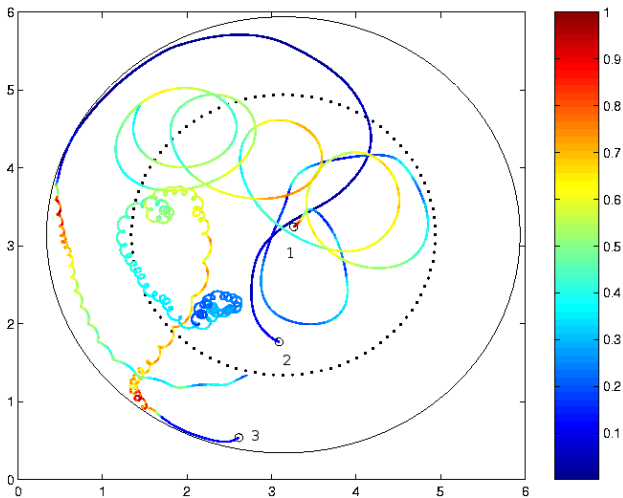}
\end{center}
\caption{Three typical trajectories in the circular geometry. The trajectories are divided into particles inside and outside the disk defined by the radius $r_0$ (circle in dotted line). Each trajectory is colored with the modulus of Lagrangian acceleration normalized by its maximum value: $|\vec a_L(t)|/max(|\vec a_L(t)|)$, where $max|\vec a_1|=3.6$, $max|\vec a_2|=11.7$ and $max|\vec a_3|=33.3$ for the particles 1, 2 and 3, respectively. The circles indicate the initial position of the particles.}
\label{Fig:COND_TRAJ_CIRCULAR}
 \end{figure}

%_______________________________________________________________________________ 
%\section{Results}
%
We now present the results of different Lagrangian quantities obtained for both geometries. We analyze for each case the Lagrangian velocity and the Lagrangian acceleration. Computations are carried out for approximately $5\cdot10^5$ timesteps, corresponding to about $403$ initial eddy turnover times. As the turbulence is freely decaying, the statistics can not be interpreted properly if the quantities considered are not made stationary. To overcome this problem we divide the Lagrangian quantities $L(t)$ by their instantaneous standard deviation computed from all particles at each time: $L(t)/\sigma_L(t)$, as suggested by Yeung \cite{Yeung_2002}. All the following statistics are studied using this normalization and for notational convenience denoted by $L(t)$. 
Three typical trajectories are shown in Fig.~\ref{Fig:COND_TRAJ_CIRCULAR}. Different behaviors can be observed: the particles can proceed in almost straight lines, spiraling motion or follow a trajectory close to the wall for a while before being reinjected into the bulk flow.
\begin{figure}%[!htb]
 \begin{center}
 \includegraphics[scale=.75]{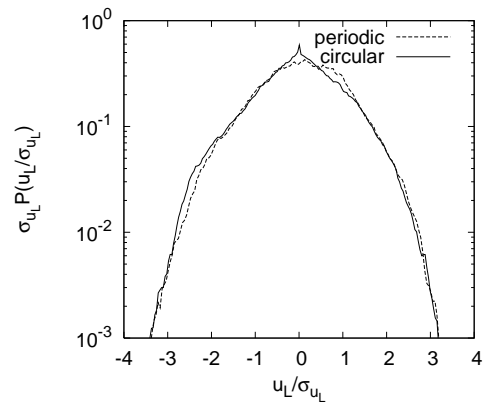}
 \end{center}
\caption{PDFs of normalized Lagrangian velocities $u_L/ \sigma_{u_L} $ where $\sigma_{u_L}= \langle u_L^2 \rangle^{1/2}$ ($\langle \cdot \rangle$ denotes the ensemble average), for the periodic geometry and for the circular geometry.}
\label{Fig:PDF_VEL}
 \end{figure}
The PDFs of the Lagrangian velocities for both geometries, shown in Fig.~\ref{Fig:PDF_VEL}, are similar and exhibit the same Gaussian-like behaviour. However, for the circular geometry, a small cusp appears around zero which indicates a large probability of values with almost zero velocity. This higher probability of velocities around zero can be explained by fluid particles that stay for relatively long times near the wall due to the no-slip boundary condition.

%%%%%%%%%%%%%%%%%%%%%%%%%%%%%%%%%%%%%%%%%%%%%%%%%%%%%%%%%%%%%%%%%%%%%%%%%%%%%%%%%%%

\begin{figure*}%[!htb]
 \begin{center}
\setlength{\unitlength}{0.5\textwidth}
\includegraphics[width=0.75	\unitlength]{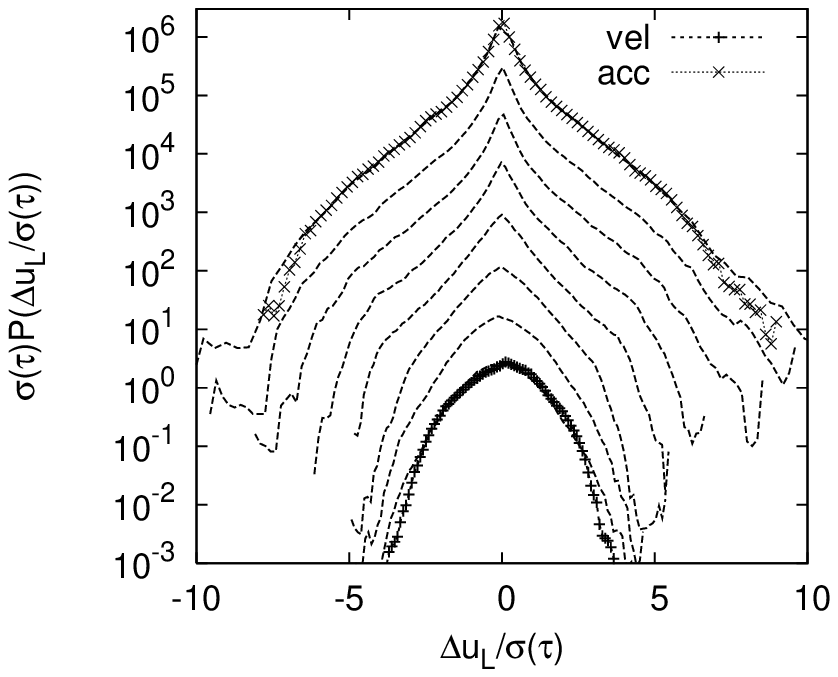}\includegraphics[width=0.75\unitlength]{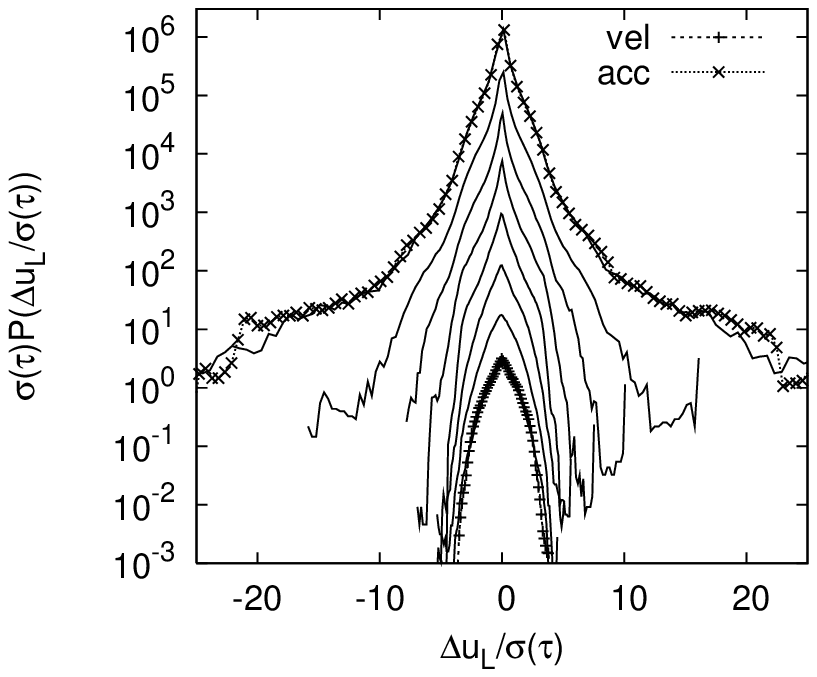}
 \end{center}
\caption{PDFs of normalized Lagrangian velocity increments $\Delta u_L(\tau) / \sigma(\tau)$ where $\sigma(\tau)= \langle (\Delta u_L(\tau))^2 \rangle^{1/2}$, for periodic (left) and circular geometry (right). The curves are shifted vertically for clarity. From top to bottom: $\tau=$ 0.1 ,0.2, 0.4, 0.8, 1.6, 3.2, 6.4, 12.8.}
\label{Fig:PDF_INCR}
 \end{figure*}

\begin{figure}%[!htb]
 \begin{center}
\setlength{\unitlength}{0.5\textwidth}
\includegraphics[width=0.75\unitlength]{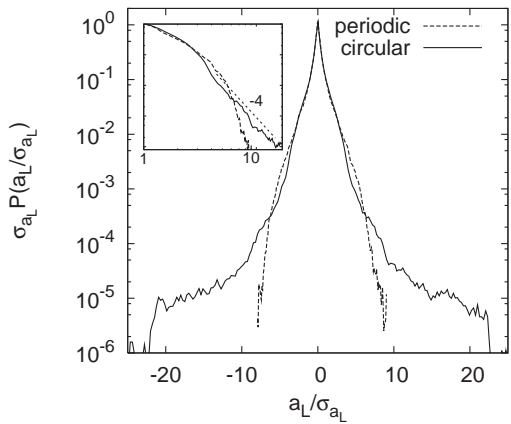}
 \end{center}
\caption{PDFs of the normalized Lagrangian acceleration $a_L/ \sigma_{a_L}$ where $\sigma_{a_L}= \langle a_L^2 \rangle^{1/2}$ for both cases. Inset: PDFs of the normalized Lagrangian acceleration in double logarithmic scale} 
\label{Fig:PDF_ACC}
 \end{figure}

In Fig.~\ref{Fig:PDF_INCR} (left) and Fig.~\ref{Fig:PDF_INCR} (right), the PDFs of the time-averaged Lagrangian velocity increments, defined by
\begin{equation}
\Delta u_L(\tau) = \langle u_L(t+\tau)-u_L(t)\rangle_t,
\end{equation}
are shown for the periodic and the confined case respectively and where $\langle \cdot \rangle_t$ denotes the time average during the entire time computation corresponding to $403T_e$. The PDFs are symmetric for both cases as is to be expected because of the symmetry of the flows. Furthermore these PDFs are qualitatively very similar to the ones obtained in experimental results in three dimensional isotropic turbulence  \cite{Mordant_PRL_2001}. For small $\tau$, the PDF of the Lagrangian velocity increments tends to the Lagrangian acceleration PDF and for large $\tau$ it tends to the PDF of the Lagrangian velocity. At small $\tau$, the PDF of the velocity increments in the circular geometry (Fig.~\ref{Fig:PDF_INCR}, right), shows
heavy tails, which are much more pronounced than the tails in the periodic case. 
This is highlighted in Fig.~\ref{Fig:PDF_ACC}, in which we superimpose the PDF of the Lagrangian acceleration for the two geometries. It is observed that the central part of the two PDFs nearly collapses. However, the tails corresponding to extreme accelerations, present a power law behavior with slope $-4$, while in the periodic case we find a stretched exponential behavior (Fig.~\ref{Fig:PDF_ACC}, inset).

\begin{figure}[!htb]
 \begin{center}
 \includegraphics[scale=0.75]{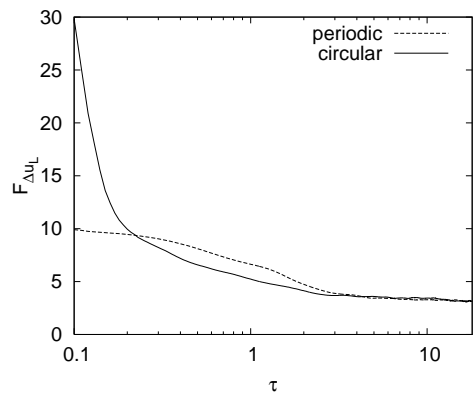}
 \end{center}
\caption{Flatness of the Lagrangian velocity increments as a function of $\tau$ for the  periodic and circular geometry.}
\label{Fig:COND_INCR}
 \end{figure}

To give a quantitative measure for the departure from Gaussianity of the Lagrangian velocity increments, its flatness is shown in Fig.~\ref{Fig:COND_INCR}.  For the Lagrangian velocity increments, at small $\tau$ the flatness tends to a value of $30$ for the circular geometry which is close to $3$ times the value of the periodic geometry. At larger $\tau$ a steep descent is observed, approaching the Gaussian value $3$ rapidly, which corresponds to the flatness of the Lagrangian velocity. In the periodic case this descent is slower.

%%%%%%%%%%%%%%%%%%%%%%%%%%%%%%%%%%%%%%%%%%%%%%%%%%%%%%%%%%%%%%%%%%%%%%%%%%%%%%%%%%

%%%%%%%%%%%%%%%%%%%%%%%%%%%%%%%%%%%%%%%%%%%%%%%%%%%%%%%%%%%%%%%%%%%%%%%%%%%%%%%%%%

\begin{figure}[!htb]
 \begin{center}
\includegraphics[scale=0.75]{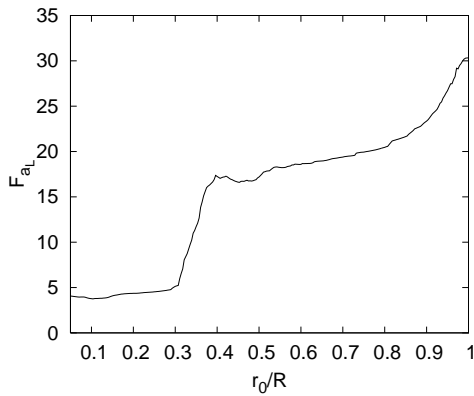}
 \end{center}
\caption{Conditional flatness of the Lagrangian acceleration as a function of radius $r_0/R$.}
\label{Fig:COND_FLAT}
 \end{figure}

From Fig.~\ref{Fig:PDF_ACC} it could be concluded that the no-slip walls are responsible for the extreme events in the acceleration. Indeed, one of the main differences between periodic and wall bounded flows, is the production of vorticity at the walls. As illustrated in Fig.~\ref{Fig:COND_TRAJ_CIRCULAR}, particles trapped in the vortices generated at the wall experience extreme acceleration, corresponding to the heavy tails of the Lagrangian acceleration PDF shown in Fig.~\ref{Fig:PDF_ACC}. These vortices are ejected from the wall by the detachment of the boundary layer, and hereby the influence of the walls can be observed in a part of the domain larger than the vicinity of the wall only. In the following we want to investigate whether these events remain confined to a region close to the wall or if the influence of the walls penetrates into the center of the domain. We proceed as follows: we choose an arbitrary radius $r_0 \le R$ and we separate the statistics into two parts, inside (denoted by $L_{r<r_0}(t)$) and outside ($L_{r>r_0}(t)$) the selected radius. A single trajectory can contribute to both regions as illustrated in Fig.~\ref{Fig:COND_TRAJ_CIRCULAR}. The flatness of the conditional Lagrangian acceleration $a_{L_{r<r_0}}(t)$ is defined as 
\begin{equation}
F_{a_L}(r)=\frac{\langle a_{L_{r<r_0}}(r) ^4\rangle}{\langle a_{L_{r<r_0}}(r) ^2\rangle^2},
\end{equation}
where $\langle \cdot \rangle$ denotes the ensemble average for the particles confined to circular subdomain defined by the radius $r_0$. It is plotted in Fig.~\ref{Fig:COND_FLAT} for the case of the circular geometry.
\\
For a radius $r_0/R<0.3$ the flatness is nearly constant with a value of about $5$. Hence, no significant influence of the wall can be found for the acceleration. When the radius is between $0.3<r_0/R<0.4$, the flatness increases rapidly, which corresponds to a sudden appearance of intermittent acceleration due to the wall. This part corresponds to a transition region between isotropic and confined turbulence. We could introduce a critical radius which measures the {\it Lagrangian boundary layer thickness $\delta_L$}, corresponding to the region $r_0/R>0.3$, in which the influence of the boundaries on the Lagrangian statistics becomes important. For $0.4<r_0/R<0.8$ the flatness increases slowly with values around 20. Finally, for $r_0/R>0.8$ the flatness strongly increases. In this region the influence of the wall becomes most important.

%%%%%%%%%%%%%%%%%%%%%%%%%%%%%%%%%%%%%%%%%%%%%%%%%%%%%%%%%%%%%%%%%%%%%%%%%%%%%%%%%%
%%%%%%%%%%%%%%%%%%%%%%%%%%%%%%%%%%%%%%%%%%%%%%%%%%%%%%%%%%%%%%%%%%%%%%%%%%%%%%%%%%
%\section{CONCLUSIONS}

To conclude, we showed by DNS of decaying two-dimensional incompressible Navier-Stokes turbulence, to what extend no-slip boundaries influence the Lagrangian statistics of velocity and acceleration. Whereas the PDF of the Lagrangian velocity is only slightly influenced by the no-slip conditions in a region close to the boundary, reflected by a small cusp around zero in its PDF, the PDF of the acceleration shows the appearance of heavy tails which are much more pronounced than in the case of periodic boundary conditions. By computing the acceleration statistics only in a subdomain of radius $r_0<R$, we were able to measure a {\it Lagrangian boundary layer thickness $\delta_L$}. For the center of the flow, outside this boundary layer, the influence on the acceleration is nearly negligible. The transition between the {\it Lagrangian boundary layer} and a region, not directly influenced by the walls, is shown to be very sharp with a sudden increase of the acceleration flatness from $\sim 5$ to $\sim 20$. Subsequently, a region of slowly increasing flatness is observed followed by a near-wall region in which, again, a sharp increase of the flatness is observed. The influence of the Reynolds number on the relation between $\delta_L$ and $r_0/R$ deserves attention and will be addressed in a more detailed study. We would like to stress the importance of the observation of a critical radius which measures a Lagrangian boundary layer thickness. In particular the fact that in our case it extends up to $r_0/R\approx 0.3$, which implies that approximately $90\%$ of the domain surface is influenced by the walls. Indeed, this information is necessary to assess the validity of the assumption of homogeneity in experimental results. If in three-dimensional turbulence $\delta_L$ is of the same order, a careful reassessment of experimental results would be needed.

%%%%%%%%%%%%%%%%%%%%%%%%%%%%%%%%%%%%%%%%%%%%%%%%%%%%%%%%%%%%%%%%%%%%%%%%%%%%%%%%%%
%\section*{Acknowledgments}

We acknowledge financial support from the Agence Nationale de la Recherche, project "M2TFP". We thank C. Baudet for fruitful discussion.
%%%%%%%%%%%%%%%%%%%%%%%%%%%%%%%%%%%%%%%%%%%%%%%%%%%%%%%%%%%%%%%%%%%%%%%%%%%%%%%%%%
%%%%%%%%%%%%%%%%%%%%%%%%%%%%%%%%%%%%%%%%%%%%%%%%%%%%%%%%%%%%%%%%%%%%%%%%%%%%%%%%%%
\vspace{-0.3cm}
\bibliographystyle{srt}

\end{document}